\documentclass[a4paper]{article}

\usepackage{spconf}
\usepackage{cite}
\usepackage{amsmath}
\usepackage[bottom]{footmisc}
\usepackage{algorithmic}
\usepackage{graphicx}
\usepackage{textcomp}
\usepackage{xcolor}
\usepackage{amsfonts}
\usepackage{amssymb}
\usepackage{epstopdf}
\usepackage{mwe}        
\usepackage{multirow}
\usepackage[export]{adjustbox}
\usepackage{tikz}
\usepackage{caption}
\usepackage{float}
\usepackage{subcaption}
\usepackage{cite}
\usepackage{algorithm}
\usepackage{kantlipsum}
\usepackage{xpatch}
\usepackage{afterpage}
\usepackage{hyperref}
\usepackage{url}
\usepackage{tablefootnote}
\usepackage{mathtools}

\DeclareMathOperator\CE{CE}

\ninept
\title{Nonparallel Emotional Voice Conversion for unseen speaker-emotion pairs using dual domain adversarial network \& Virtual Domain Pairing}
\name{Nirmesh Shah$^{1*}$, Mayank Singh$^{1*}$, Naoya Takahashi$^2$,  Naoyuki Onoe $^{1}$\thanks{$^*$ both authors have equal contribution}}
\address{$^1$Sony Research India,
  $^2$Sony Group Corporation, Japan\\ Email:\{Nirmesh.Shah,Mayank.A.Singh,Naoya.Takahashi,Naoyuki.Onoe\}@sony.com\vspace{-0.5cm}}

\begin{document}

\maketitle
\begin{abstract}
Primary goal of an emotional voice conversion (EVC) system is to convert the emotion of a given speech signal from one style to another style without modifying the linguistic content of the signal. Most of the state-of-the-art approaches convert emotions for seen speaker-emotion combinations only. In this paper, we tackle the problem of converting the emotion of speakers whose only neutral data are present during the time of training and testing (i.e., unseen speaker-emotion combinations). To this end, we extend a recently proposed StartGANv2-VC architecture by utilizing dual encoders for learning the speaker and emotion style embeddings separately along with dual domain source classifiers. For achieving the conversion to unseen speaker-emotion combinations, we propose a Virtual Domain Pairing (VDP) training strategy, which virtually incorporates the speaker-emotion pairs that are not present in the real data without compromising the min-max game of a discriminator and generator in adversarial training. We evaluate the proposed method using a Hindi emotional database.\\
%
\end{abstract}

\section{Introduction}
Emotional voice conversion (EVC) attempts to modify perceived emotion style in a given speech signal to a particular target emotion style without modifying the linguistic content of the speech signal \cite{zhou2022emotional}. Early stage of EVC approaches \cite{tao2006prosody,aihara2014exemplar,luo2016emotional,ming2016deep,gao2018nonparallel,robinson2019sequence} mostly focused on speaker-dependent scenarios where they converted emotion styles for a single speaker. Recently, several works attempt to alter both perceived speaker identity and emotion style of a given source speech signal \cite{kishida2020simultaneous,du2021identity}. These methods have potential applications in various tasks such as movie dubbing, conversational assistance, cross-lingual synthesis, etc. Most of the previous approaches can convert the emotion of a speaker whose emotional data is present either at the time of training \cite{moritani2021stargan,he2021improved,zhou2021limited,choi2020stargan} or testing \cite{qian2021global,zhou2021seen}.
However, collecting emotional voice for target speakers is often expensive, time-consuming, and sometimes impossible. In this paper, we address the problem of converting the emotion of speakers whose only neutral data (data having only neutral emotion) is present by leveraging emotional speech data from other supporting speakers. In particular, the target speakers' emotional voice data is never available during the training and testing, and the proposed model learns to infer emotional voice of the target speakers from other supporting speakers of whom we have emotional data.
For brevity, we call speaker-emotion pairs present in the dataset as seen pairs and pairs having speaker-emotion combinations absent in the dataset as unseen pairs.

Some EVC approaches tackle unseen emotion cases by putting aside an entire category of emotion style during training \cite{zhou2021seen,schnell2021emocat,qian2021global}. \cite{zhou2021seen} achieves the EVC task for an unseen emotion by utilizing a pre-trained emotion classifier for generating emotional embeddings. However, their focus is to convert a voice to a particular unseen emotion class for a speaker who has emotional data for several seen emotion categories. In contrast, our aim is to convert emotion for speakers who do not have any emotional recordings. Some approaches attempt emotion voice conversion for the unseen speakers, i.e., speaker-independent EVC \cite{zhou2020converting,shankar2019multi}, by putting aside a few speakers as a heldout database \cite{zhou2020converting,shankar2019multi}. However, these methods utilize both neutral and emotional data from the seen speakers during training. These approaches do not include data from speakers whose only neutral data is available during the training. In addition, these methods are one-to-one and they cannot convert simultaneously both speaker identity as well as emotional identity. However, our approach is many-to-many and can convert both speaker and emotion identities together. To the best of the authors' knowledge, this is the first attempt to consider the specific case of the EVC task for unseen speaker-emotion pair, which fits in the more realistic dubbing applications of EVC, since we often have neutral data but not the emotional data from the desired target speaker. 
\begin{figure*}[htb]
    \centering
    \includegraphics[width=0.85\linewidth]{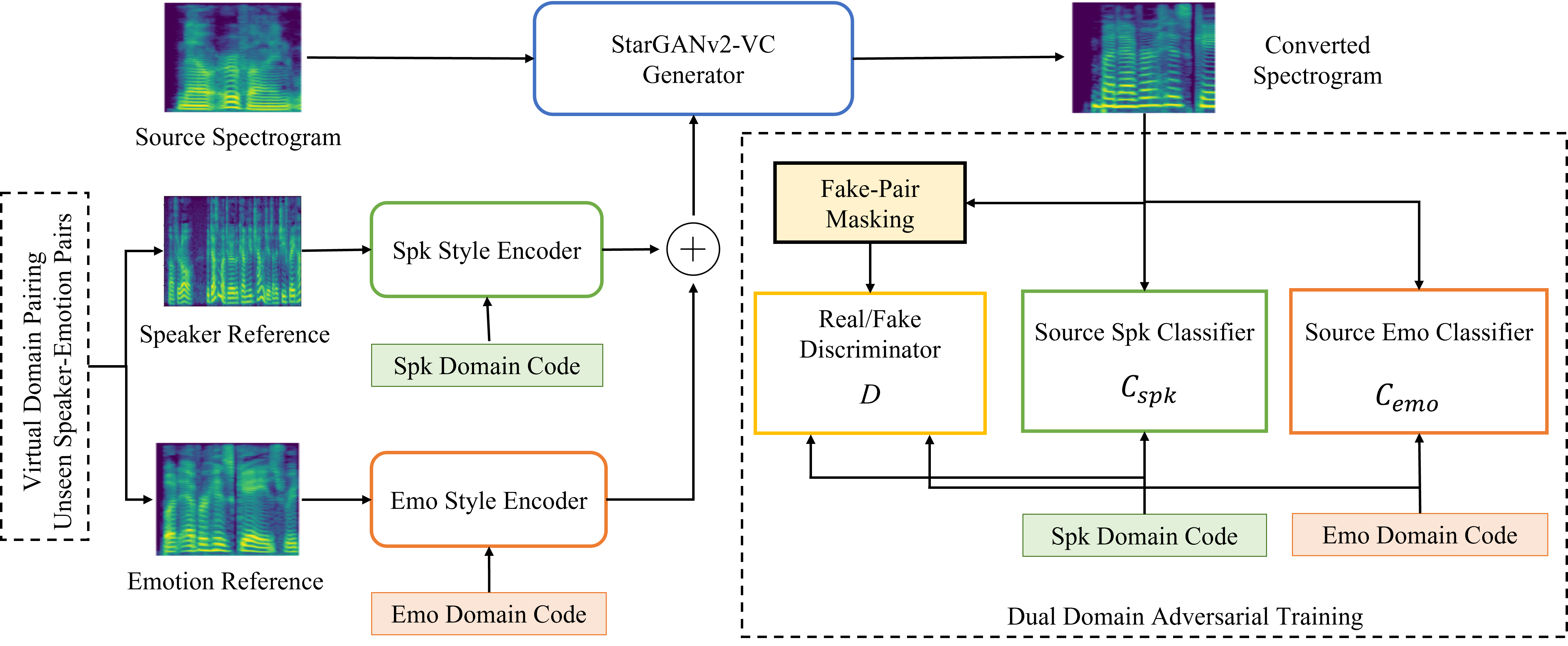}
    \caption{Block diagram of the proposed EVC-USEP architecture.}
    \label{fig:evc_block}
\vspace{-0.2cm}
\end{figure*}

\indent Among various approaches for the EVC task, autoencoder \cite{gao2018nonparallel,zhou2020converting,schnell2021emocat,qian2021global} and generative adversarial network (GAN)-based approaches \cite{zhou2021seen,he2021improved,starganv2vc} are capable of achieving emotional voice conversion in the case of non-parallel training data. 
Recently, a StarGANv2-based \cite{choi2020stargan} approach called StarGANv2-VC has demonstrated excellent effectiveness over different style conversion tasks \cite{starganv2vc}. However, the original StarGANv2-VC is designed to separately perform either speaker conversion for seen target speakers or EVC for seen emotion style and a seen target speaker. In this paper, we first modify the StarGANv2-VC architecture for converting the speaker and emotion styles simultaneously in a unified model by utilizing two encoders for learning speaker style and emotion style embeddings along with dual domain source classifiers for classifying source speaker and the emotion style. We then devise training strategies to achieve EVC for \textit{U}nseen \textit{S}peaker-\textit{E}motion \textit{P}airs (i.e., EVC-USEP) by using emotional data from supporting speakers. In particular, we propose a Virtual Domain Pairing (VDP) training strategy, which randomly generates the combinations of speaker-emotion pairs that are not present in the real data without compromising the min-max game of a discriminator and generator in adversarial training. In particular, a fake-pair masking (FPM) strategy is proposed to ensure that the discriminator does not overfit because of the fake pairs. We refer our proposed system as EVC-USEP throughout the paper.

We experimentally show that EVC-USEP successfully converts the emotion of speakers who do not contain any emotional speech during training and testing. We present the effectiveness of the proposed model against baseline and several ablation studies via objective and subjective evaluations. The key contributions from the paper are summarized below.
{\setlength{\leftmargini}{12pt} 
\begin{itemize}
\item We propose EVC for the Unseen speaker-emotion Pairs (EVC-USEP) task, where target speakers do not have any emotional data during both training and testing time.
    \item We propose an EVC-USEP network by incorporating two separate encoders and dual domain source classifiers.
    \item We further propose Virtual Domain Pairing (VDP) and fake-pair masking (FPM) training strategies for generalizing the model to unseen speaker-emotion pairs.
    \item We have also applied the annealing strategy in our proposed architecture which improves the emotion conversion accuracy.
\end{itemize}
}
\section{Related work}
Broadly, EVC approaches can be classified into two categories, parallel and non-parallel approaches. In parallel EVC task, the training data consists of utterances having the same linguistic content that is spoken by the speakers in multiple emotions. Although parallel datasets allow the usage of supervised learning-based methods which are in general easier to train, one of the disadvantages of using such datasets is the unnaturalness of having uncorrelated emotion and linguistic content. Also, collecting parallel dataset is expensive and sometimes not possible. Earlier studies have applied Gaussian mixture model (GMM) \cite{tao2006prosody}, sparse representations \cite{aihara2014exemplar}, deep belief network \cite{luo2016emotional}, sequence-to-sequence \cite{ming2016deep,zhou2021limited,robinson2019sequence} and rule-based approaches \cite{xue2018voice} to achieve parallel EVC task.

In the case of non-parallel EVC, the utterances may not have the same linguistic content across multiple speakers. This allows the utterances to have correlated linguistic content and emotion style. Unsupervised methods, namely, CycleGAN \cite{liu2020emotional,zhou2020transforming,shankar2020non}, and autoencoder-based approaches \cite{gao2018nonparallel,zhou2020converting,schnell2021emocat,qian2021global} have been proposed for non-parallel EVC tasks. These methods are suitable for seen speaker-emotion pairs. Recently, variational autoencoding Wasserstein generative adversarial network (VAW-GAN) based methods have been explored that can be extended to achieve EVC for unseen speakers and unseen emotion cases \cite{zhou2020converting,zhou2021seen}. Apart from this, StarGAN \cite{moritani2021stargan,he2021improved} and StarGANv2-VC \cite{starganv2vc} approaches have also become popular for speaker dependent EVC task. 
\section{Proposed method}
\subsection{EVC-USEP}
The proposed EVC-USEP architecture is shown in Fig. \ref{fig:evc_block}. It extends StarGANv2-VC \cite{starganv2vc} by treating the speaker style and emotion styles as individual domains (as shown in Fig. \ref{fig:evc_block}). To this end, we incorporate two style encoders, $S_{sp}$ and $S_{em}$, that separately extract the speaker style embedding $h_{sp}=S_{sp}(R_{sp},y_{sp})$ and the emotion style embedding $h_{em}=S_{em}(R_{em},y_{em})$ from reference spectrograms $R_{sp}$, $R_{em}$ to specify the target speaker and emotion style, respectively. Here, $y_{sp}$ is the speaker domain code and $y_{em}$ is the emotion domain code. The domain codes are used to apply domain specific projections to obtain the style embeddings.


The generator $G$ converts the style of source spectrogram $X$ into the target style specified by the style embeddings as $G(X,h_{f0},h_{sp},\\h_{em})$. 
As in the original StarGANv2-VC, we also incorporate the information of the fundamental frequency by conditioning the generator with a pitch embedding $h_{f0}$ extracted by a pre-trained joint detection and classification (JDC) network \cite{jdc} to guide the conversion. Hereafter, we omit $h_{f0}$ for clarity.
Along with the style encoders, we also train two mapping networks for generating speaker and emotion style embeddings from a latent variable sampled from a normal distribution, as done in StarGANv2-VC for the single domain case \cite{starganv2vc}. This enables us to specify the target style without having any references during the inference time. 
For adversarial training, we use a domain-specific discriminator $D(.,y)$ which predicts whether a sample is real or converted (fake). Here, $y$ is the concatenation of $(y_{sp},y_{em})$. Also, $y_{src}$ and $y_{trg}$ are for source and target speakers, respectively. It has shared layers for all the domains, followed by a domain-specific linear layer. In addition, we incorporate two separate domain classifiers $C_{sp}$ and $C_{em}$ which classify the source speaker domain and emotion domain of the converted samples, respectively.




We use the following training objectives for optimization.\\
\textbf{Adversarial loss:} 
The generator $G$ learns to generate a realistic mel sepctrogram via the domain-specific adversarial loss:
\begin{equation}
\label{eq:adv}
    \begin{split}
        \mathcal{L}_{adv}= &
        \mathbb{E}_{X,y_{src}}[\log D(X,{y_{src}})] + \\ & 
        \mathbb{E}_{X,y_{trg},h_{sp},h_{em}}[\log (1-D(G(X,h_{sp},h_{em}),y_{trg}))].
    \end{split}
\end{equation}
\textbf{Adversarial source classifier loss:} The two source domain classifiers, $C_{sp}, C_{em}$ are trained to classify the source speaker and emotion, respectively, using cross entropy loss $\CE(.)$:
\begin{equation}
\label{eq:advcls}
    \begin{split}
        \mathcal{L}_{advcls}=&
        \mathbb{E}_{X,y_{sp},h_{sp}}[\CE(C_{sp}(G(X,h_{sp},h_{em})),y_{sp})] + \\ &
        \mathbb{E}_{X,y_{em},h_{sp}}[\CE(C_{em}(G(X,h_{sp},h_{em})),y_{em})].
    \end{split}
\end{equation}
where, $y_{sp}$ and $y_{em}$ are from source domain codes. In contrast, the generator is trained to minimize $\mathcal{L}_{advcls}$ by taking $y_{sp}$ and $y_{em}$ from target domain codes.\\
\textbf{Style reconstruction loss:} Style reconstruction loss is used to ensure that the style codes can be reconstructed from the converted sample.
\begin{equation}
\label{eq:sty}
    \begin{split}
       \mathcal{L}_{sty}=&\mathbb{E}_{X,y_{sp},h_{sp}}|| h_{sp}-S_{sp}(G(X,h_{sp},h_{em}),y_{sp})||_1\\
       & \mathbb{E}_{X,y_{em},h_{em}}|| h_{em}-S_{em}(G(X,h_{sp},h_{em}),y_{em})||_1
    \end{split},
\end{equation}
where, $y_{sp}$ and $y_{em}$ are from reference domain codes.\\
\textbf{Style diversification loss:} To ensure the diversity of generated samples with different style embeddings, we modified a style diversification loss to incorporate both speaker and emotion styles. We extract two sets of style embeddings, $(h_{sp},h'_{sp})$ and $(h_{em}, h'_{em})$, from the same domain $y_{sp},y_{em}$ and maximize the L1 distance of samples generated using different combination of style embeddings as
\begin{equation}
\label{eq:ds}
    \begin{split}
        &\mathcal{L}_{ds}=\mathbb{E}_{X,y_{sp},h_{sp}}|| G(X,h_{{sp}},h_{{em}})-G(X,h_{sp},h'_{em})||_1\\
        &+\mathbb{E}_{X,y_{sp},h_{sp}}|| G(X,h_{sp},h_{em})-G(X,h'_{sp},h_{em})||_1\\
        &+\mathbb{E}_{X,y_{sp},h_{sp}}|| G(X,h'_{sp},h_{em})-G(X,h'_{sp},h'_{em})||_1.
    \end{split}
\end{equation}
\textbf{$F_0$ consistency loss}: To produce $F_0$ consistent converted voices, $F_0$ consistent loss is utilized and it is given by \cite{starganv2vc}:
\begin{equation}
    \mathcal{L}_{F0}=\mathbb{E}_{X,h_{sp},h_{em}}[||\hat{F}(X)-\hat{F}(G(X,h_{sp},h_{em}))||_1],
\end{equation}
where $\hat{F}(X)$ provides normalized $F_0$ values from an input mel spectrogram.\\
\textbf{Norm consistency loss:} We utilize norm consistency loss ($\mathcal{L}_{norm}$) in order to preserve speech/silence interval in converted spectrum and it is given by \cite{starganv2vc}:
\begin{equation}
    \mathcal{L}_{norm}=\mathbb{E}_{X,h_{sp},h_{em}}\Bigg[\frac{1}{T}\sum_{t=1}^{T}\Big|  ||X_{.,t}||- ||G(X,h_{sp},h_{em})_{.,t}|| \Big|\Bigg],
\end{equation}
where $||X_{.,t}||$ is norm of absolute column sum for a mel spectorgram $X$ with total $T$ frames at the $t^{th}$ frame. We propose modifications to $\mathcal{L}_{F0}$ and $\mathcal{L}_{norm}$, which is discussed in sub-section 3.4. Other losses, such as, the speech consistency loss and cycle consistency loss have been preserved as it is from StarGANv2-VC\cite{starganv2vc}. Finally, the full objective is the weighted sum of all these losses.

\subsection{Virtual Domain Pairing (VDP)}
At the time of inference, we aim at converting the emotion style of the source speaker's voice by providing an emotion style embedding of the target emotion. However, since emotional data is available only for the supporting speakers but not for the target speakers, the model does not generalize well to unseen emotion of target speakers if we train the model with speaker-emotion pairs which exist in the training data. Fortunately, the proposed EVC-USEP architecture enables us to virtually incorporate speaker-emotion pairs that do not exist in the training data. When we sample style embeddings in Eq. \eqref{eq:adv}, \eqref{eq:advcls}, \eqref{eq:sty}, and \eqref{eq:ds}, we independently sample the target speaker domain $y_{sp}$ and the emotion domain $y_{em}$ and compute style embeddings by using either the style encoders or the mapping networks.
This allows us to virtually generate emotional references for target speakers having only neutral data by pairing them with emotional references from supporting speakers. We refer to the proposed sampling strategy as Virtual Domain Pairing (VDP). 

\subsection{Fake-Pair Masking (FPM)}
When using the VDP, the discriminator is always trained to predict the fake class for unseen speaker-emotion pairs because for unseen speaker-emotion case, real data is unavailable. Therefore, even if the generated samples sound realistic, the discriminator will collapse, and always predict the converted samples as fake.
In such a case, the generator can try to fool the discriminator by generating a voice that sounds like a seen pair, which limits the conversion capability for unseen speaker-emotion pairs and can promote undesired speaker conversion.
To address this problem, we introduce the FPM strategy. We mask samples of unseen speaker-emotion pairs for the real-fake discriminator training and thus, the discriminator is trained only on mel-spectrograms from seen speaker-emotion pairs. In contrast, the goal of source classifiers is to predict the source domain regardless of real or fake pairs. Thus, domain classifiers are only responsible for converting the emotion and speaker identity of unseen pairs. Hence, they should be less prone to fake pairs and should not suffer from the aforementioned problem. In particular, We have experimentally shown the effectiveness of FPM and VDP strategies by doing the ablation studies reported in section 4. 
\subsection{Annealing strategy}
$F_0$ consistency and norm consistency losses ensure that normalized $F_0$ value and speech/silence interval remain same after emotion conversion. However, we hypothesize that $F_0$ and speech/silence regions are emotion dependent and as such should be modified during emotion conversion. Hence, we applied a loss annealing strategy (i.e., we linearly decrease the weights assigned to these losses) with each epoch while training the proposed EVC-USEP model. In addition, we have also trained one model w/o annealing (i.e., given in \cite{starganv2vc}) for comparison and one model w/o $\mathcal{L}_{F0}$ and $\mathcal{L}_{norm}$ in order to check the effectiveness of the annealing strategy on these losses.   
\section{Experiments}
\subsection{Experimental Setup}
\textbf{Hindi Emotional Speech Database:} To demonstrate the ability of our models to modify the emotion of speakers having only neutral speech data, we train our models on a set of target speakers having only neutral voices and a set of supporting speakers having multiple emotions. We have utilized an internally developed 54 hours of Hindi emotional speech corpus for this task. The dataset is recorded by total \textit{9} speakers. We have collected text from stories and categorized these texts in six emotions, namely, Neutral, Happy, Sad, Fear, Surprise, and Angry. Each speaker has recorded 1 hour of data in each emotion. In particular, we have utilized emotional data from \textit{7} speakers and only neutral data from remaining \textit{2} speakers for training with an aim to test the EVC models for the unseen speaker-emotion cases. The Hindi database is downsampled to 24kHz. We randomly split all the utterances in the dataset for training and testing in the ratio of 9:1. \\
\textbf{Model Details:} We train all the networks for 150 epochs, with a batch size of 10. The speaker and emotion domain classifiers are introduced in the training process after 50 epochs. Both source domain classifiers have the same architecture as the real-fake discriminator, which follows the same architecture given in StarGANv2-VC \cite{starganv2vc}. The annealing loss strategy is applied after 50 epochs by decreasing the weightage of the $\mathcal{L}_{F0}$, \& $\mathcal{L}_{norm}$ losses linearly from 5 to 0. The other training details are kept the same as reported in \cite{starganv2vc}.
During inference, we convert the emotion of speech samples taken from 2 speakers, whose only neutral data was utilized during training. 
We use a pre-trained Parallel WaveGAN\cite{pwg} to synthesize waveforms from mel-spectrograms as in StarGANv2-VC\cite{starganv2vc}. The mapping network is used for extracting style embedding during inference.\\ 
\textbf{Baseline Model and Ablation Studies}: To the best of authors` knowledge this is the first attempt to consider the specific case of the EVC task for unseen speaker-emotion pair given only neutral emotional data of the speaker. Among earlier approaches, VAW-GAN-based EVC was proposed to tackle the case of unseen speaker \cite{zhou2020converting}. Thus, for baseline, we have considered training the VAW-GAN-based EVC model while utilizing unseen speaker-emotion pair data \cite{zhou2020converting}. However, the VAW-GAN-based EVC model is limited to converting just the emotion whereas our model can convert the emotion as well as the speaker identity. We evaluate our model on the more challenging task of converting both emotion and speaker identity, but because of the limitations of VAW-GAN we have kept the same source and target speakers during conversion for the case of baseline.
In addition to the comparison with the baseline model, we conducted several ablation studies to verify the effectiveness of the VDP and FPM strategies. For this, we do an ablation study on the VDP and FPM. Moreover, the proposed EVC-USEP architecture has utilized annealing strategies for $\mathcal{L}_{F0}$ and $\mathcal{L}_{norm}$. Hence, we do an ablation study on the annealing strategy by comparing proposed architecture w/o annealing case and w/o $\mathcal{L}_{F0}$ and $\mathcal{L}_{norm}$ loss cases.
\begin{table}[t]
\caption{Subjective and objective evaluations results. MOS are shown for quality along with margin of error corresponding to 95\% confidence interval.}
\vspace{-0.3cm}
\label{table:1}
\centering
\resizebox{0.5\textwidth}{!}{%
\begin{tabular}{cccc}
\hline
      & Subjective  & \multicolumn{2}{c}{Objective}           \\
\textbf{Method}                                                                                  & \textbf{Quality}        & \textbf{Spk. SIM}   & \textbf{Emo. Acc}   \\ \hline \hline
VAW-GAN \cite{zhou2020converting} & 1.90 $\pm$ 0.02  & 0.61 & 34.03\\ 
EVC-USEP (Proposed) & \textbf{4.21 $\pm$ 0.01} & \textbf{0.76}  & \textbf{41.50}  \\ \hline
w/o VDP & 3.82 $\pm$  0.02 & 0.72 & 24.83   \\
w/o FPM & 3.92 $\pm$ 0.02 & 0.74 & 37.76  \\
w/o annealing & 3.96 $\pm$ 0.02 & 0.73 & 36.73 \\
w/o $\mathcal{L}_{F0}$ and $\mathcal{L}_{norm}$ & 3.87 $\pm$ 0.02             & 0.71               & 30.95              \\ \hline
\end{tabular}
}
\vspace{-0.3cm}
\end{table}
\begin{figure}[t]
    \centering
    \includegraphics[width=1.0\linewidth]{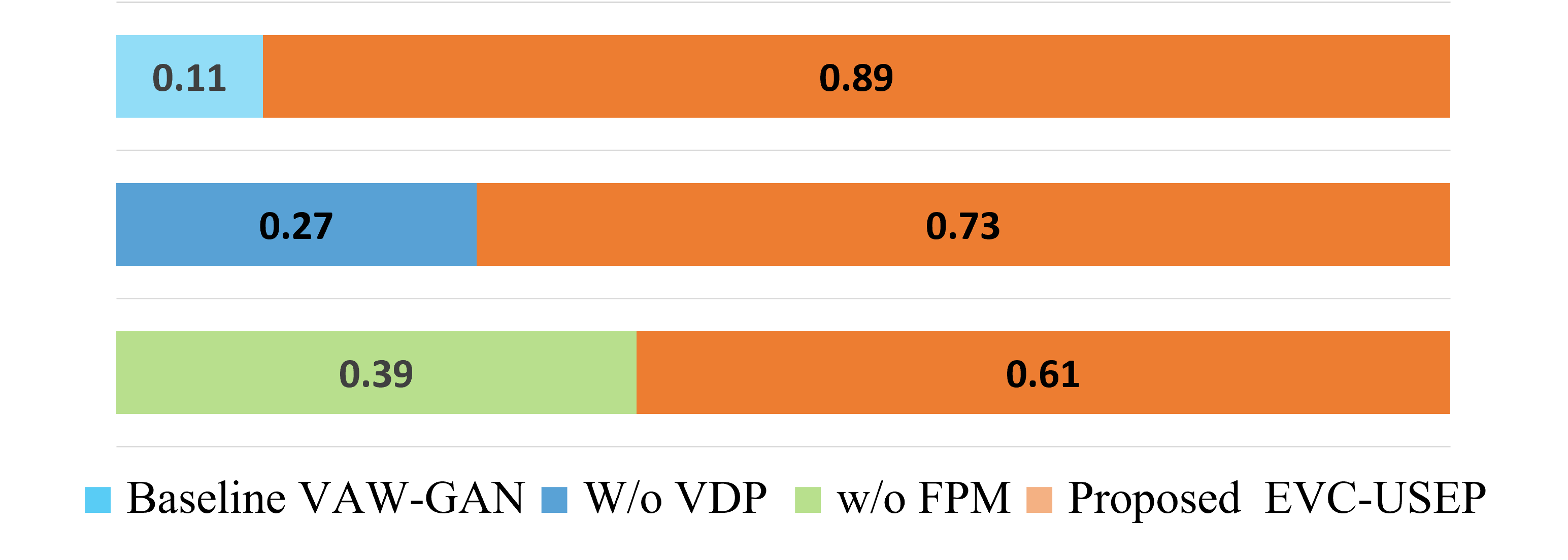}
    \center
    \vspace{-0.6cm}
        \caption{ABX Subjective Evaluation for Emotion Similarity.}
    \label{fig:abx}
    \vspace{-0.6cm}
\end{figure}
\vspace{-0.5cm}
\subsection{Subjective Evaluation}
We have conducted two subjective tests, namely, mean opinion scores (MOS) and ABX test to evaluate the quality of converted voices and evaluation of emotion conversion, respectively. Demo audio samples can be found online\footnote{Available at \href{https://demosamplesites.github.io/EVCUP/}{https://demosamplesites.github.io/EVCUP/}}. Total 14 evaluators (with no known hearing impairments and having age between 18 to 34 years) participated in both subjective tests. We asked evaluators to rate the naturalness of each audio clip on a scale of 1 to 5, where, 1 indicates completely distorted quality and 5 indicates high-quality natural voice. A total of 112 samples are evaluated for each system in the MOS test. The results of the subjective evaluation, as shown in Table \ref{table:1}, conclude that our model outperforms the baseline model in terms of the quality of converted voices. The ablation study also shows that the proposed VDP and FPM strategies are crucial for achieving high-quality emotion conversion for speakers having only neutral data. Moreover, the tests confirm that our proposed idea of applying the annealing strategy to $\mathcal{L}_{F0}$ and $\mathcal{L}_{norm}$ improves the results compared to w/o annealing and w/o $\mathcal{L}_{F0}$ and $\mathcal{L}_{norm}$ experiments. 

\indent To evaluate the emotion conversion, we conducted three ABX tests between 1) baseline VAW-GAN and proposed EVC-USEP 2) w/o VDP model and proposed EVC-USEP and 3) w/o FPM model and proposed EVC-USEP. Given two converted samples, we asked the evaluators to prefer the converted voice which has more similar emotion w.r.t. the provided target emotional reference audio signal. A total of 140 samples are considered for evaluating each system in the ABX test. Result of ABX tests are shown in Fig. \ref{fig:abx}. We can observe that the proposed model has preferred more in the context of emotion conversion w.r.t. the baseline. In addition, the proposed model is preferred more compared to the system w/o VDP and w/o FPM in the context of emotional similarity.
\subsection{Objective Evaluation}
Since our goal is to achieve EVC for speakers having only neutral data, we do not have ground truth for the converted samples. Hence, deterministic evaluation metrics such as spectral distortion-based objectives are not appropriate for the problem we consider. Alternatively, we use an emotion classification network to evaluate the accuracy of emotion conversion. We train the emotion classification model on the Hindi emotional dataset in a similar way to \cite{zhou2021seen}.  
Classification accuracy obtained on the converted samples are summarized in Table \ref{table:1}. From Table \ref{table:1}, we observe that accuracy of our proposed model is significantly higher compared to the baseline. The ablation study again confirms the effectiveness of VDP and FPM strategies for unseen speaker-emotion pairs.
The speaker similarity score is used to evaluate speaker similarity of the converted voices. We calculate it by taking the inner product between the speaker embeddings extracted from the converted samples and a reference target speaker's sample \cite{wan2018generalized,Resemblyzer}. Table \ref{table:1} also contains speaker similarity scores for different experiments and we can observe that our proposed model has higher speaker similarity scores. 
\section{Conclusion}
We proposed the EVC-USEP architecture to tackle the problem of converting the emotion of target speakers whose only neutral data is present by leveraging emotional speech data from other supporting speakers. We further propose Virtual Domain Pairing and FPM strategies which are shown to be essential for achieving the emotional voice conversion for unseen speaker-emotion pairs task.
From both objective and subjective evaluations, we confirm that the proposed method successfully converts the emotion of the target speakers, outperforming the baselines w.r.t. emotion similarity, speaker similarity, and quality of the converted voices, while achieving decent naturalness. In the future, we would like to further improvise the speaker conversion capability of the proposed model for the unseen target speaker case.
\bibliographystyle{IEEEtran.bst}
\bibliography{mybib.bib}
\end{document}